\documentclass[a4paper,10pt,aps,prl,twocolumn,superscriptaddress,showpacs,amsmath,amssymb]{revtex4}
\usepackage{graphicx,color}

\newcommand{\dd}{\mathrm{d}}

\newcommand{\pd}[2]{\frac{\partial #1}{\partial #2}}

\newcommand{\mean}[1]{\langle #1 \rangle}

\newcommand{\IInt}[3]{\int_{#2}^{#3}\dd #1\;}

\renewcommand{\vec}[1]{\mathbf #1}

\DeclareMathOperator{\sech}{sech}

\newcommand{\gam}{\gamma}

\newcommand{\kap}{\kappa}

\newcommand{\im}{\text{i}}

\newcommand{\kT}{k_\text{B}T}
\newcommand{\nois}{\boldsymbol\xi}
\newcommand{\x}{\vec r}

\newcommand{\tx}{\tau_\text{r}}

\newcommand{\lp}{\ell_\text{p}}
\newcommand{\pii}{p^\text{(i)}}
\newcommand{\paa}{p^\text{(a)}}
\newcommand{\pt}{p_\text{T}}
\newcommand{\pn}{p_\text{N}}

\begin{document}

\title{Negative interfacial tension in phase-separated active suspensions}

\author{Julian Bialk\'e}
\author{Hartmut L\"owen}
\affiliation{Institut f\"ur Theoretische Physik II,
  Heinrich-Heine-Universit\"at, D-40225 D\"usseldorf, Germany}
\author{Thomas Speck}
\affiliation{Institut f\"ur Physik, Johannes Gutenberg-Universit\"at Mainz,
  Staudingerweg 7-9, 55128 Mainz, Germany}

\begin{abstract}
  We study numerically a model for active suspensions of self-propelled
  repulsive particles, for which a stable phase separation into a dilute and a
  dense phase is observed. We exploit that for non-square boxes a stable
  ``slab'' configuration is reached, in which interfaces align with the
  shorter box edge. Evaluating a recent proposal for an intensive active
  swimming pressure, we demonstrate that the excess stress within the
  interface separating both phases is negative. The occurrence of a negative
  tension together with stable phase separation is a genuine non-equilibrium
  effect that is rationalized in terms of a positive stiffness, the estimate
  of which agrees excellently with the numerical data. Our results challenge
  effective thermodynamic descriptions and mappings of active suspensions onto
  passive pair potentials with attractions.
\end{abstract}


\maketitle


Equilibrium statistical physics~\cite{chandler} rests on two deceptively
simple premises: the laws of conservation and the uniform probability of all
accessible microstates in isolated systems. Of course, suitable local
equilibria are only a small part of the universe, and non-equilibrium
encompasses so many diverse processes and phenomena that the quest for a
universal description is one of the great challenges in statistical
physics. While likely futile in full generality, there are subclasses of
driven systems for which a comprehensive theory seems to be in reach. One such
class are suspensions of active particles.

Active matter~\cite{roma12,marc13,elge14} has emerged as a paradigm to
describe a broad wealth of non-equilibrium collective, dynamical behavior
ranging from droplets~\cite{thut11} to bacteria~\cite{wens12} down to
microtubule networks driven by molecular motors~\cite{sanc12}. Here we focus
on suspensions of self-propelled colloidal spherical particles suspended in a
solvent (see Ref.~\citenum{bial14} for a short perspective of these systems
and references) or polymer solution~\cite{line12}. Quite strikingly, particles
cluster into dense and dilute regions for high enough density and swimming
speeds. Such a behavior has been observed both
experimentally~\cite{theu12,pala13,butt13} and in computer simulations of
purely repulsive
particles~\cite{bial13,redn13,fily14,sten13,sten14,wyso14,taka14}. It is
understood microscopically to arise from the time-scale separation between the
decorrelation time of the directed motion and the collision rate, which is
controlled by speed and density. The actual time-scales depend on many details
(pair potentials, swimming mechanisms, hydrodynamic
interactions~\cite{zott14}) but the generic effect is robust and only requires
volume exclusion in combination with a persistent motion of the particles.

Since the formation and growth of dense domains indeed resembles the phase
separation of passive suspensions with attractive interactions, several
theoretical descriptions following a ``thermodynamical'' route have been
proposed: effective mean-field free energies~\cite{tail08,spec14,cate14},
pressure equations of state~\cite{witt14,taka14,taka14a,gino14}, and mappings
to effective isotropic pair potentials~\cite{line12,das14}. However,
microscopic interactions of the self-propelled particles are not isotropic and
the crucial physical ingredient, as mentioned, is the persistence of motion
over a length $\lp=v_0\tx$, where $v_0$ is the swimming speed and $\tx$ the
time over which orientations decorrelate. In this Letter, we numerically test
the idea of an \emph{intensive} pressure in active suspensions assuming an
equation of state exists~\cite{solo14}. We adopt a strategy that has proven to
be very fruitful in the study of phase-separated passive systems by exploiting
finite-size transitions in non-square simulation
boxes~\cite{schm14}. Following old ideas by Kirkwood and Buff~\cite{kirk49}
together with a generalization of the swimming pressure~\cite{taka14} gives us
access to the interface~\cite{evan79}, and we show that the interfacial
\emph{tension} is actually negative. In contrast, the \emph{stiffness}
governing the interface fluctuations is positive, and we show how to relate
both through the dissipated work.


We simulate a minimal model for active particles that has been studied by a
range of groups~\cite{bial13,redn13,fily14,sten13,sten14,wyso14,taka14}. The
model consists of $N$ particles with diameter $a$ interacting via short-ranged
repulsive forces (here from a Weeks-Chandler-Andersen potential $u(r)$, for
details and parameters see Refs.~\citenum{butt13,bial14}). The dynamics is
overdamped,
\begin{equation}
  \dot\x_i = -\nabla_i U + v_0\vec e_i + \nois_i,
\end{equation}
where $\nois_i$ is the Gaussian translational noise with zero mean and
correlations $\mean{\nois_i(t)\nois_j^T(t)}=2\delta_{ij}\mathbf
1\delta(t-t')$, and $U=\sum_{j<i}u(|\x_i-\x_j|)$ is the total potential
energy. We consider the two dimensional case with a simulation box of size
$L_x\times L_y$ employing either periodic boundary conditions, or walls in the
$x$ direction and periodic boundaries in the $y$ direction. Every particle
swims with fixed speed $v_0$ along its unity orientation $\vec e_i$, which
undergoes free rotational diffusion with diffusion coefficient $1/\tx$. We
employ dimensionless quantities such that lengths are measured in units of $a$
and time in units of $D_0/a^2$, where $D_0$ is the bare translational
diffusion coefficient. The no-slip boundary condition then implies
$\tx=\tfrac{1}{3}$. Moreover, energies are measured in units of $\kT$ for
fixed solvent temperature $T$.


\begin{figure}[t]
  \centering
  \includegraphics[width=\linewidth]{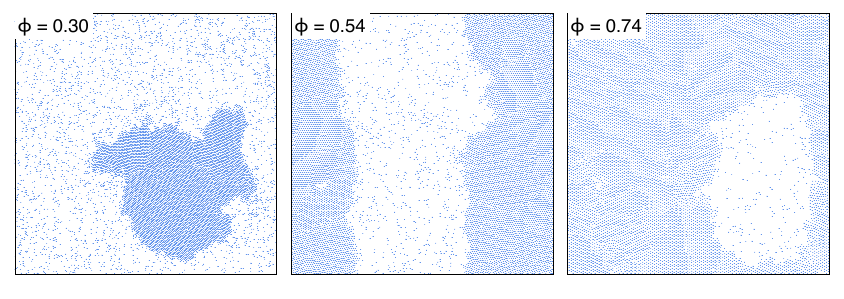}
  \caption{Finite-size transitions of active Brownian particles (in a box with
    aspect ratio 1.2) going from low to high density: droplet, slab, and
    bubble.}
  \label{fig:scan}
\end{figure}

\begin{figure}[t]
  \centering
  \includegraphics{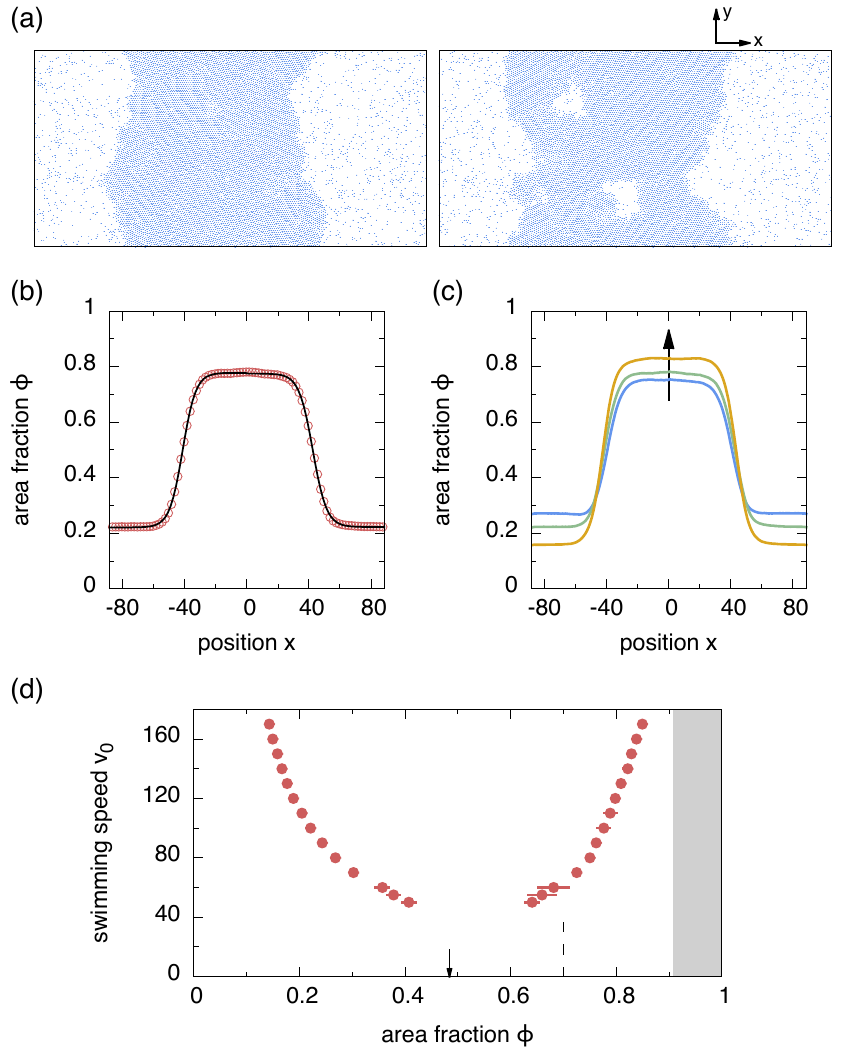}
  \caption{Slab geometry: (a)~Snapshots of a phase separated system with
    aspect ratio $L_x/L_y =2$. A dense slab is surrounded by the dilute gas
    phase. Large fluctuations occur, not only at the interface but also the
    dense inner region of the slab might develop ``holes'' (right
    snapshot). (b)~Measured density profile for $v_0=100$ (symbols) and fit of
    Eq.~\eqref{eq:fit} (line). (c)~Measured density profiles for
    $v_0=80,100,150$ (from bottom to top) from which we extract the coexisting
    densities. (d)~Resulting phase diagram: The symbols show the coexisting
    densities $\phi_\pm$ with errors estimated from 5 independent runs (except
    $v_0=170$ with two runs). The dashed vertical line indicates the freezing
    area fraction $\phi\simeq0.7$ and the shaded area the excluded packing
    fractions for real hard disks. The arrow indicates the global density
    $\phi=0.49$ at which simulations are run.}
  \label{fig:slabs}
\end{figure}

We first scan the system for swimming speed $v_0=180$ and vary the global
density $\bar\rho=N/A$. As shown in Fig.~\ref{fig:scan}, we observe
finite-size transitions as we increase the density: from the homogeneous
suspension to a droplet of the dense phase, to a slab, to a ``bubble'' (or
void) forming within the dense phase. These transitions appear to be exact
counterparts of the transitions observed in simulations of vapor-liquid
coexistence in finite volumes~\cite{schr09}. While the snapshots in
Figs.~\ref{fig:scan} and~\ref{fig:slabs}(a) show a high degree of local order
in the dense phase, these crystalline patches have only a short lifetime and
constantly reorganize. Hence, particles do not freeze and the description as
an active liquid-vapor coexistence is more appropriate.

To make comparisons with passive suspensions easier, densities will be
reported as area fractions $\phi=\bar\rho\pi(a^\ast/2)^2$ using an effective
hard-sphere diameter $a^\ast=0.984a$ obtained via Barker-Henderson from the
pair potential~\cite{bark67}. Such a mapping is known to work well for passive
repulsive suspensions although at high swimming speeds it will certainly
become less reliable. In the following, we exploit the slab configuration and
all simulations are run at $\phi=0.49$ with $N=10,000$ particles varying the
speed $v_0$. In analogy to simulations of passive fluids, we employ a
non-square box of area $A=L_xL_y$ with edge lengths $L_x>L_y$ such that the
slab of the dense phase is encouraged to span the shorter length, see
Fig.~\ref{fig:slabs}(a). At high enough swimming speeds $v_0$, such slabs form
spontaneously and remain stable. In order to reach the steady state faster,
all $N$ particles are initially placed in a dense slab in the middle of the
system. After a relaxation time of $t_\text{rel}=100$ we start to collect and
analyze data.

Qualitatively, looking at the simulations one notes that fluctuations are much
more violent than expected from a passive suspension. In particular, even in
the dense phase larger ``bubbles'' might form, see
Fig.~\ref{fig:slabs}(a). Still, given sufficient statistics, the averaged
density profiles excellently fit the mean-field functional form
\begin{equation}
  \label{eq:fit}
  \phi(x) = \frac{\phi_++\phi_-}{2} + \frac{\phi_+-\phi_-}{2}
  \tanh\left(\frac{x-x_0}{2w}\right),
\end{equation}
see Fig.~\ref{fig:slabs}(b). Here, $x_0$ marks the midpoint of the profile and
$w$ is related to the width of the interface. Density profiles are measured
from the simulations by dividing the simulation box into slices with area
$A_1$, where $x$ is the distance of the slice from the
center-of-mass. Although the two interfaces are correlated, in a first attempt
we treat them independently and perform separate fits for $x<0$ and $x>0$. The
interfacial width $w$ and bulk phase densities $\phi_\pm$ are then obtained by
taking the mean of the results for the left and right half of the
box. Measured density profiles for several speeds are shown in
Fig.~\ref{fig:slabs}(c). For each profile, we fit Eq.~\eqref{eq:fit} from
which we extract the coexisting densities $\phi_\pm$ shown in
Fig.~\ref{fig:slabs}(d). Note that the error increases as we go to lower
speeds as expected from critical fluctuations.


We now study the mechanical stress generated in the active suspension. To this
end, we focus on a single swimming speed $v_0=100$. Note that the system is
translationally invariant in the $y$-direction since we have encouraged the
slab to align that way. Clearly, phase separation and the occurrence of
interfaces breaks the translational invariance in $x$-direction so that
averaged quantities can only depend on $x$. The condition of hydrostatic
equilibrium $\nabla\cdot\vec p=0$ then implies that the total pressure $\vec
p(x)$ is a diagonal tensor and, moreover, that the normal pressure
$p_{xx}=\pn$ is constant throughout the box to ensure mechanical stability. In
contrast, the tangential pressure $p_{yy}(x)=\pt(x)$ can, and does, vary
spatially with $x$.

\begin{figure}[t]
  \centering
  \includegraphics{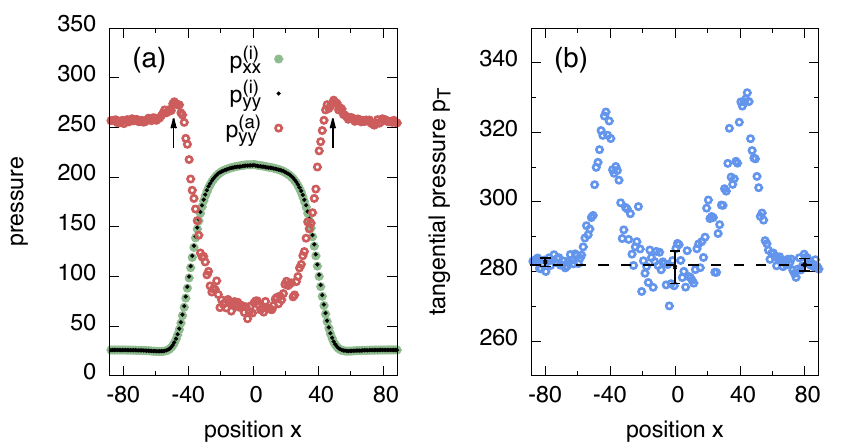}
  \caption{Pressure profiles for $v_0=100$: (a)~The diagonal components
    $\pii_{xx}$ ($\bullet$) and $\pii_{yy}$ ($\cdot$) of the interaction
    pressure (which lie on top of each other) and the tangential active
    pressure $\paa_{yy}$ ($\circ$). Note the increase of the active pressure
    in the interface before it drops in the dense phase (arrows). (b)~The
    total tangential pressure $\pt=\rho+\pii_{yy}+\paa_{yy}$. The dashed line
    is the estimate for the bulk pressure $p\simeq282$, the error bars show
    the root-mean-square errors of local horizontal fits.}
  \label{fig:press}
\end{figure}

We first consider the pressure tensor
\begin{equation}
  \label{eq:pi}
  \vec\pii(x) = \frac{1}{2A_1}\mean{\x_{ij}\vec f_{ij}^T}_x
\end{equation}
due to particle interactions, where $\x_{ij}=\x_i-\x_j$ is the connecting
vector of particles $i$ and $j$, and $\vec f_{ij}$ is the pair force along
this vector due to the repulsive potential. The brackets $\mean{\cdot}_x$
denote the average over particle pairs for which at least one particle is
within the slice at $x$. The factor $\tfrac{1}{2}$ has to be included to
compensate for the fact that every bond crossing between slices is counted
twice. Note that there are alternative spatial discretization schemes, all of
which lead to the same integrated pressure~\cite{walt83}. The two diagonal
components $\pii_{xx}$ and $\pii_{yy}$ are plotted in
Fig.~\ref{fig:press}(a). Both curves lie on top of each other and follow
qualitatively the density, \emph{i.e.}, the interaction pressure is low in the
dilute phase and high in the dense phase. Clearly, there is something missing
since such an inhomogeneous pressure is mechanically unstable and violates the
condition of hydrostatic equilibrium.

Only very recently, the idea that due to their directed motion the particles
exert a mechanical stress has been formalized by Brady and
coworkers~\cite{taka14}. Following their approach, the scalar \emph{active
  pressure} can be calculated via
\begin{equation}
  \label{eq:pa:glo}
  \paa = \frac{v_0}{2A}\sum_{i=1}^N\mean{\vec e_i\cdot\x_i},
\end{equation}
where $\x_i$ is indeed the absolute position. The active pressure thus stems
from the correlations between particle positions and orientations. Assuming a
gas of non-interacting swimmers with $\dot\x_i=v_0\vec e_i+\nois_i$, we
obtain~\cite{taka14}
\begin{equation}
  \label{eq:pa:id}
  \paa_\text{id} = \frac{v_0}{2A}\sum_{i=1}^N\IInt{t'}{-\infty}{t}
  \mean{\vec e_i(t)\cdot\dot\x_i(t')} = \frac{1}{2}\bar\rho v_0^2\tx
\end{equation}
using the correlation function $\mean{\vec e(t)\cdot\vec e(t')}=e^{-|t-t'|/\tx}$.

To consider the spatial dependence of the active pressure~\eqref{eq:pa:glo},
we introduce the generalized tensor
\begin{equation}
  \label{eq:pa}
  \vec\paa(x) = \frac{v_0}{A_1}\mean{\vec e_i\x_i^T}_x
\end{equation}
in analogy to Eq.~\eqref{eq:pi}. The average is now taken over the subset of
particles that at time $t$ occupy slice $x$. However, there is a subtlety here
since this destroys the correlations between the $x$-coordinate and the
orientations, which, as Eq.~\eqref{eq:pa:id} demonstrates, depend not only on
the configuration but on the previous history. Hence, only the component
$\paa_{yy}(x)$ is actually meaningful, which is plotted in
Fig.~\ref{fig:press}(a). It again qualitatively follows the density but is now
inverted with respect to the interaction pressure: the active pressure is high
in the dilute region and drops considerably in the dense region. The physical
reason is that particle motion is hindered in the dense phase and orientation
and actual displacement are thus less correlated.

Two conceptual insights into the nature of active suspensions are gained by
plotting the total tangential pressure
$\pt(x)=\rho(x)+\pii_{yy}(x)+\paa_{yy}(x)$ (there is also the ideal gas
contribution $\rho(x)$, which, however, is small). As demonstrated in
Fig.~\ref{fig:press}(b), the bulk pressures of dense and dilute phase are
equal, which in turn implies $\pn\approx\pt$. To corroborate that normal and
tangential bulk pressure coincide, we have studied walls allowing to directly
measure $\pn$ as the mechanical pressure exerted onto the walls~[SM]. The
first insight is thus that the swimming pressure of Takatori \emph{et al.} is
indeed the missing link to define and measure a pressure that is
\emph{intensive}. What is quite striking is that the pressure within the
interface is larger than the bulk pressure. Identifying the interfacial
tension with the excess stress (the factor $\tfrac{1}{2}$ again accounts for
the two interfaces) leads to~\cite{kirk49}
\begin{equation}
  \label{eq:gam}
  \gam = \frac{1}{2}\IInt{x}{0}{L_x}[\pn-\pt(x)] \simeq -842,
\end{equation}
which becomes negative. This is the second, quite surprising insight. While it
has no consequence for the mechanical stability, our intuition tells us that a
system with a negative tension cannot be stable. The reason is that in systems
for which classical thermodynamics is applicable, the interfacial tension
determines the excess free energy due to the presence of interfaces. A
negative tension implies that the suspension could lower its free energy by
creating more interfaces, leading again to a homogeneous state. Quite in
contrast, in active suspensions one observes a stable, phase-separated state.

Note that the excess stress is entirely due to the active pressure since
$\pii_{xx}=\pii_{yy}$, which means in particular that there is no energetic
contribution from the potential energy. Indeed, in Fig.~\ref{fig:press}(a) we
observe an increase of the swimming pressure entering the interface before it
drops. This can be understood qualitatively: the instantaneous interface has a
small width [cf. snapshots Fig.~\ref{fig:slabs}(a)] and acts like a (flexible)
wall. Swimmers accumulate but are still (comparably) free to slide along the
interface in the $y$ direction, and hence their larger density (with respect
to the dilute region) leads to a higher active tangential pressure
$\paa_{yy}$. Another puzzling observation is the magnitude of $|\gam|$, which
is huge compared to typical values $\sim1$ in passive liquids (\emph{e.g.},
for vapor-liquid coexistence in the Lennard-Jones fluid in two dimensions
$\gam_\text{LJ}\simeq0.42$ has been reported~\cite{sant08}).


\begin{figure}[t]
  \centering
  \includegraphics{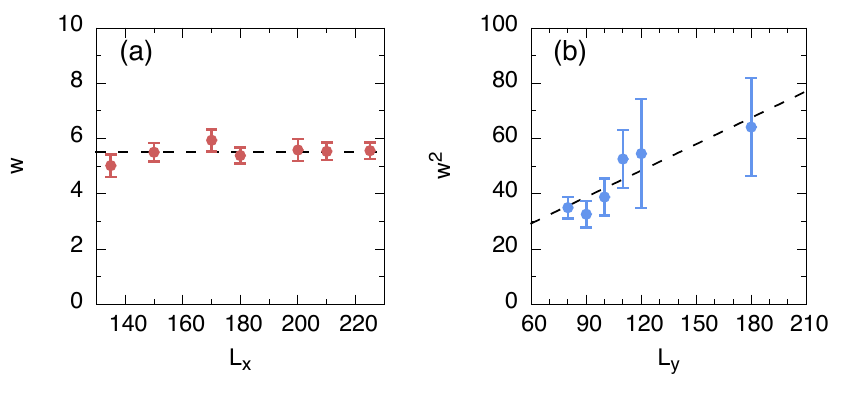}
  \caption{Interfacial width $w$ for $v_0=100$: (a)~As a function of box
    length $L_x$ for constant $L_y=90$. The dashed line is a constant
    fit. (b)~As a function of $L_y$ for constant $L_x=200$. The dashed line is
    a fit of Eq.~\eqref{eq:cwt} with $w_0\simeq3.18$ and
    $\kap\simeq0.26$. Error bars show standard deviation of 10 independent
    runs.}
  \label{cwt}
\end{figure}

To reach a better understanding, we now study the interfacial width $w$ in
more detail. Fig.~\ref{cwt} shows $w$ obtained from several simulation runs at
speed $v_0=100$ through fitting Eq.~\eqref{eq:fit}. We systematically study
different system sizes by holding one box length fixed and varying the
other. The total number of particles varies such that the global density is
kept constant for all data points. While changing $L_x$ does not influence the
width, we observe an increase of $w$ when increasing $L_y$. This behavior
demonstrates two things: First, the system sizes considered here are large
enough to have reached a constant width as we vary $L_x$. Second, the
dependance on $L_y$ agrees with standard capillary wave theory (CWT) assuming
equipartition. Hence, it is instructive to recall the arguments leading to
CWT~\cite{evan79}: One assumes an ideal instantaneous interface, in our case a
line of total length $\ell$, which separates the two phases. To change this
length, work has to be spent against the positive line tension. Assuming no
overhangs, one can decompose the profile $h(y)=\sum_q h_qe^{\im qy}$ into
Fourier modes $h_q$. Since the energy for every mode stems from the thermal
environment, equipartition implies $\mean{|h_q|^2}=(L_y\kap q^2)^{-1}$, where
$\kap$ is the interfacial stiffness governing the fluctuations. For passive
liquid-vapor coexistence, this stiffness is equal to the tension (in units of
$\kT$ per unit length).

To estimate the interfacial width $w$, we calculate the fluctuations of the
instantaneous interface~[SM],
\begin{equation}
  \label{eq:cwt}
  w^2 \approx \sum_q \mean{|h_q|^2} = w_0^2 + \frac{L_y}{12\kap},
\end{equation}
which predict a linear divergence due to the capillary waves. The offset
$w_0^2$ corresponds to fluctuations of the $q=0$ mode, which are bounded due
to the periodic boundary conditions. Moreover, we have assumed that even in
the driven active suspension equipartition holds. While the use of
equipartition is of course not rigorous, the predicted leading linear
dependence on $L_y$ agrees quite well with the simulation data in
Fig.~\ref{cwt}(b). It can be further motivated by the fact that orientational
degrees of freedom do not develop long-ranged correlations (even in the
phase-separated case). Using Eq.~\eqref{eq:cwt} we can thus fit the data in
Fig.~\ref{cwt}(b) to extract the stiffness $\kap\simeq 0.26$, which is both
positive and small. That it is positive agrees with the observation of stable
phase separation and finite-size transitions, that it is small agrees
qualitatively with the observed strong fluctuations.

Finally, to rationalize a positive stiffness with a negative tension, recall
that every particle swims with fixed velocity $v_0$, \emph{i.e.}, from the
particle's perspective it pumps the surrounding fluid against its own
hydrodynamic drag. Hence, the particles constantly spent a ``housekeeping''
work $\mathcal W<0$ on the solvent. The typical scale of this work per
particle is the hydrodynamic force times the persistence length,
$\mathcal{W}/N=-v_0\lp=-v_0^2\tx$ (in Ref.~\citenum{taka14a} this expression
appears as a positive energy scale). As long as the work per length $\gam$
gained from extending the interface is smaller, the interface is stable. The
housekeeping work thus fulfills a role similar to the thermal energy in
passive suspension. For the stiffness we then find
$\kap\approx\gam/(-v_0\lp)\simeq0.25$ for $v_0=100$, which compares favorably
with the value extracted from the fitted interfacial widths.


In summary, we have demonstrated that the mechanical interfacial tension in
phase-separated active suspensions is negative. This implies that work is
released when the interfacial length $\ell$ is increased. However, this work
is not ``available'' to the suspension but part of the work that is spent by
the particles to drive the surrounding fluid. We expect that a negative
tension is not specific to the model studied here but holds more generally in
active matter. In principle, it can be observed in particle-resolved
experiments~\cite{theu12,pala13,butt13} with a stabilized interface. The
incorporation of both a negative tension and correct interfacial fluctuations
into thermodynamic descriptions based on an effective free energy, a concept
that seems to work well for the \emph{bulk} phases~\cite{cate14,taka14a}, is
certainly challenging. The deeper reason is that in thermal equilibrium the
same free energy determines the probability of fluctuations away from typical
configurations, a connection that no longer holds for systems driven away from
thermal equilibrium.


\acknowledgments

We thank J\"urgen Horbach, Peter Virnau, and Kurt Binder for helpful
discussions and comments. We gratefully acknowledge financial support by DFG
within priority program SPP 1726 (grant numbers SP 1382/3-1 and LO 418/17-1).



\newpage

\section{Interface fluctuations}

\begin{figure}[b!]
  \centering
  \includegraphics{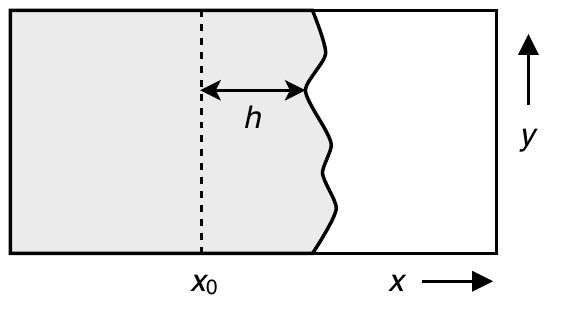}
  \caption{Sketch of the instantaneous interface, where $h(y)$ is the distance
    from the midpoint $x_0$ of the averaged profile.}
  \label{fig:iface}
\end{figure}

For completeness, here we provide more detailed informations regarding the
determination of the interface fluctuations. We consider a single interface
with midpoint $x_0=0$ and we assume that we could determine an instantaneous
interface in form of a line such that $h(y)$ denotes the position as a
function of $y$, see Fig.~\ref{fig:iface}. We decompose the profile into
Fourier modes
\begin{equation*}
  h(y) = \sum_q h_q e^{\im qy} \quad\text{with}\quad
  h_q = \frac{1}{L_y}\IInt{y}{0}{L_y} h(y) e^{-\im qy}
\end{equation*}
and determine the interfacial width due to fluctuations through
\begin{equation}
  \label{eq:w}
  w^2 = \frac{1}{L_y}\IInt{y}{0}{L_y} \mean{[h(y)]^2} 
  = \sum_q \mean{|h_q|^2}.
\end{equation}

\subsection{Equipartition}

In passive suspensions in thermal equilibrium, the excess (free) energy due to
the interface is $E_\text{s}=\gam\ell$, where $\gam$ is the interfacial
tension and $\ell$ is the length of the interface. Expanding to lowest order
in the gradient, one finds
\begin{equation*}
  \ell = \IInt{y}{0}{L_y} \sqrt{1+[h'(y)]^2}
  \approx L_y + \frac{1}{2}L_y\sum_q q^2|h_q|^2
\end{equation*}
which is quadratic in the Fourier coefficients. Hence, equipartition implies
\begin{equation*}
  \mean{|h_q|^2} = \frac{1}{\kap L_y q^2}
\end{equation*}
with $\kap=\gam$. Plugging this relation back into Eq.~\eqref{eq:w}, we obtain
\begin{equation}
  \label{eq:cwt}
  w^2 = w_0^2 + \frac{2}{\kap L_y} \sum_{q>0} \frac{1}{q^2} 
  = w_0^2 + \frac{L_y}{12\kap},
\end{equation}
where $w_0^2$ describes the fluctuations of the $q=0$ mode and the second term
the contribution due to the undulations (capillary waves) of the interface
line. For this result we have employed $q=\frac{2\pi}{L_y}n$ due to the
periodic boundaries together with the sum
\begin{equation*}
  \sum_{n=1}^\infty \frac{1}{n^2} = \frac{\pi^2}{6}.
\end{equation*}
For the active suspension we assume that Eq.~\eqref{eq:cwt} still holds albeit
now with a stiffness $\kap\neq\gam$.

\subsection{Density profile}

From the profile $h(y)$ we can construct the instantaneous density profile
\begin{equation*}
  \hat\rho(x) = \frac{1}{L_y}\IInt{y}{0}{L_y} \left[ 
    \rho_+\theta(x-h(y)) + \rho_-\theta(h(y)-x) \right],
\end{equation*}
where $\theta(x)$ is the unit step (Heaviside) function. The derivative of the
mean profile thus reads
\begin{equation*}
  \pd{\mean{\hat\rho}}{x} = (\rho_+-\rho_-)\mean{\delta(x-h(y))}.
\end{equation*}
Due to translational invariance, the expectation value becomes independent of
$y$. It can be calculated from the characteristic function
\begin{equation*}
  \mean{e^{\im kh}} = \exp\left\{ -2k^2\sum_{q>0}\frac{1}{\kap L_y q^2}
  \right\} = e^{-k^2w^2}
\end{equation*}
again assuming equipartition. Performing the reverse transformation, we obtain
\begin{equation*}
  \pd{\mean{\hat\rho}}{x} \propto \frac{\rho_+-\rho_-}{w}
  \exp\left\{-\frac{x^2}{4w^2} \right\}
\end{equation*}
for the variation of the mean density profile.

In the main text, we have chosen to fit density profiles using the mean-field
expression
\begin{equation}
  \label{eq:rho}
  \rho(x) = \frac{\rho_++\rho_-}{2} + \frac{\rho_+-\rho_-}{2}
  \tanh\left(\frac{x}{2w}\right).
\end{equation}
The spatial derivative reads
\begin{equation*}
  \pd{\rho}{x} = \frac{\rho_+-\rho_-}{2w}\sech^2\left(\frac{x}{2w}\right)
  \approx \frac{\rho_+-\rho_-}{2w} \exp\left\{-\frac{x^2}{4w^2} \right\},
\end{equation*}
where $\sech x=1/\cosh x$. For the final result, we have expanded $\ln\sech
x\approx-x^2/2$ to second order. This demonstrates that we can estimate the
width $w$ appearing in Eq.~\eqref{eq:rho} using Eq.~\eqref{eq:cwt}.

\section{Walls}

\begin{figure}[t]
  \centering
  \includegraphics{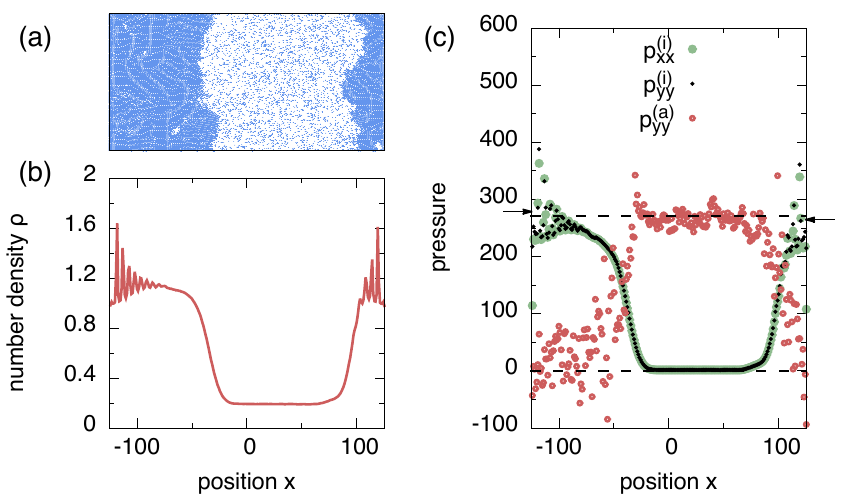}
  \caption{Phase-separated active suspension in the presence of walls:
    (a)~Snapshot and (b)~density profile showing the strong layering close to
    the walls. (c)~The different pressures as a function of $x$ (the distance
    from the box center). The arrows indicate the pressure exerted onto the
    walls measured independently for left and right wall, and the upper dashed
    line shows their mean value $\simeq272$. The lower dashed line indicates
    zero pressure.}
  \label{fig:walls}
\end{figure}

We have also studied the active suspension in the presence of walls in the $x$
direction. As shown in Fig.~\ref{fig:walls}(a), the particles now accumulate
close to the walls and leave a dilute region between. We again determine the
density profile, however, now the absolute distance $x$ from the center of the
simulation box is used. Although we now use $N=20,000$ particles to improve
statistics, obtaining good data is far more difficult compared to periodic
boundaries. The main reason is the strong layering of particles close to the
wall, see Fig.~\ref{fig:walls}(b).

The walls consist of a short-ranged potential. The advantage is that we can
now determine the force exerted on the walls, and therefore the pressure
$\pn$, directly. As demonstrated in Fig.~\ref{fig:walls}(c), within errors we
find the same value as expected for an intensive pressure. Moreover, it agrees
with the sum of active and interaction pressure in the dilute region,
$\pt\approx\pn$. Note that the interaction pressure in the dilute phase
becomes close to zero, whereas the swimming pressure close to the walls
(apparently) averages to zero as well.

\end{document}